\documentclass[sigconf,nonacm]{acmart}
\usepackage{booktabs}
\usepackage{multirow}
\usepackage{xspace}
\setcopyright{none}
\newcommand{\nItems}{10,307\xspace}
\newcommand{\nEvalItems}{8,800\xspace}
\newcommand{\nRefProbes}{1,507\xspace}
\newcommand{\hcN}{60\xspace}

\newcommand{\hcNscored}{57\xspace}
\newcommand{\hcAgreeN}{47\xspace}
\newcommand{\hcAgreePct}{82.5\%\xspace}
\newcommand{\hcAgreeWPct}{89.5\%\xspace}
\newcommand{\hcKappa}{0.64\xspace}
\newcommand{\hcTypeKappa}{0.26\xspace}
\newcommand{\hcDisagree}{10\xspace}

\newcommand{\hcFinalTypeDiff}{3\xspace}

\newcommand{\hcFrameN}{263\xspace}

\newcommand{\hcCauseRendered}{6\xspace}
\newcommand{\hcCauseRedirect}{2\xspace}
\newcommand{\hcCauseForeign}{1\xspace}
\newcommand{\hcCausePlaceholder}{1\xspace}

\newcommand{\fetchStart}{2026-10-19\xspace}
\newcommand{\fetchEnd}{2026-10-21\xspace}
\newcommand{\fetchDays}{three\xspace}
\newcommand{\nSitesS}{1,507\xspace}
\newcommand{\sitesSone}{300\xspace}
\newcommand{\pagesSone}{595\xspace}
\newcommand{\unusableSonePct}{9.8\%\xspace}
\newcommand{\probeSonePct}{10.5\%\xspace}
\newcommand{\sitesStwo}{400\xspace}
\newcommand{\pagesStwo}{631\xspace}
\newcommand{\unusableStwoPct}{26.3\%\xspace}
\newcommand{\probeStwoPct}{13.3\%\xspace}
\newcommand{\sitesSthree}{107\xspace}
\newcommand{\pagesSthree}{186\xspace}

\newcommand{\probeSthreePct}{17.3\%\xspace}
\newcommand{\sitesSfour}{400\xspace}
\newcommand{\pagesSfour}{622\xspace}

\newcommand{\probeSfourPct}{13.7\%\xspace}
\newcommand{\sitesSfive}{300\xspace}
\newcommand{\pagesSfive}{464\xspace}

\newcommand{\probeSfivePct}{13.1\%\xspace}
\newcommand{\sitesP}{3,383\xspace}
\newcommand{\pagesP}{4,795\xspace}
\newcommand{\unusablePPct}{23.2\%\xspace}
\newcommand{\nPurls}{4,795\xspace}
\newcommand{\nPairs}{6,156\xspace}
\newcommand{\nPages}{7,293\xspace}
\newcommand{\kOneTwoHundredPct}{60.2\%\xspace}
\newcommand{\unusableAllPct}{20.9\%\xspace}
\newcommand{\unusableLoPct}{20.0\%\xspace}
\newcommand{\unusableHiPct}{23.5\%\xspace}
\newcommand{\autoTestPagesPct}{25.7\%\xspace}
\newcommand{\nTestMdpi}{36\xspace}
\newcommand{\nTestMdpiUnusable}{35\xspace}
\newcommand{\autoTestNoMdpiPct}{22.7\%\xspace}
\newcommand{\probeAllPct}{13.1\%\xspace}
\newcommand{\blockedAllPct}{21.5\%\xspace}
\newcommand{\blockedPPct}{25.7\%\xspace}
\newcommand{\kTwoRescuePct}{43.5\%\xspace}
\newcommand{\PaccessPct}{8.2\%\xspace}
\newcommand{\PaccessTopFour}{scribd.com, mdpi.com, biomedcentral.com, springer.com\xspace}
\newcommand{\PaccessTopFourPct}{67\%\xspace}
\newcommand{\shellRescuePct}{75\%\xspace}
\newcommand{\kTwoNotAPct}{16.7\%\xspace}
\newcommand{\goldUnusablePct}{20.0\%\xspace}
\newcommand{\goldUnusableCI}{13.3--27.7\%\xspace}
\newcommand{\autoUnusableTestPct}{25.5\%\xspace}
\newcommand{\nGold}{300\xspace}
\newcommand{\nGoldUnsure}{6\xspace}
\newcommand{\autoRecall}{0.96\xspace}
\newcommand{\autoPrecision}{0.75\xspace}
\newcommand{\autoTypeAcc}{81\%\xspace}
\newcommand{\autoTypeKappa}{0.65\xspace}

\newcommand{\nBrowser}{50\xspace}
\newcommand{\browserN}{47\xspace}
\newcommand{\browserMiss}{3\xspace}
\newcommand{\browserMissPct}{6.4\%\xspace}
\newcommand{\hbestGoldRecallPct}{86\%\xspace}

\newcommand{\recHone}{0.25\xspace}

\newcommand{\recHtwo}{0.76\xspace}

\newcommand{\recHbest}{0.86\xspace}
\newcommand{\faHbest}{0.056\xspace}
\newcommand{\autoRecHbest}{0.74\xspace}
\newcommand{\autoFaHbest}{0.02\xspace}

\newcommand{\recMone}{0.57\xspace}

\newcommand{\recMonetypeD}{0.33\xspace}

\newcommand{\recMtwo}{0.91\xspace}
\newcommand{\faMtwo}{0.133\xspace}
\newcommand{\autoRecMtwo}{0.78\xspace}
\newcommand{\autoFaMtwo}{0.05\xspace}

\newcommand{\faMtwoP}{0.27\xspace}
\newcommand{\recMthree}{0.82\xspace}
\newcommand{\faMthree}{0.133\xspace}

\newcommand{\oracleMtwo}{0.69\xspace}
\newcommand{\oracleMthree}{0.76\xspace}
\newcommand{\oracleHbest}{0.84\xspace}
\newcommand{\goldNun}{99\xspace}
\newcommand{\goldNus}{195\xspace}
\newcommand{\goldTypeC}{12\xspace}
\newcommand{\nHbesttypeC}{11\xspace}
\newcommand{\goldTypeD}{48\xspace}
\newcommand{\nHbesttypeD}{39\xspace}
\newcommand{\goldTypeE}{30\xspace}
\newcommand{\nHbesttypeE}{27\xspace}
\newcommand{\moneTypeAcc}{77\%\xspace}

\newcommand{\primDiff}{+0.051\xspace}
\newcommand{\primDiffCI}{[+0.011, +0.100]\xspace}
\newcommand{\usableShiftPts}{12.2\xspace}
\newcommand{\driftRatio}{3.8\xspace}
\newcommand{\driftHzero}{21.9\%\xspace}
\newcommand{\liveUsableHzero}{52.6\%\xspace}
\newcommand{\driftBaseHzero}{1,254\xspace}
\newcommand{\driftAuto}{5.8\%\xspace}
\newcommand{\liveUsableAuto}{40.5\%\xspace}
\newcommand{\driftBaseAuto}{981\xspace}
\newcommand{\driftMone}{11.4\%\xspace}

\newcommand{\driftHbest}{10.2\%\xspace}

\newcommand{\nDriftHzero}{274\xspace}
\newcommand{\nDriftUnusable}{217\xspace}
\newcommand{\driftUnusableShare}{79\%\xspace}
\newcommand{\nDriftAbsent}{96\xspace}
\newcommand{\nDriftAccess}{73\xspace}
\newcommand{\nDriftDegraded}{48\xspace}
\newcommand{\deadP}{11.0\%\xspace}
\newcommand{\failedP}{36.4\%\xspace}

\newcommand{\testDriftMtwo}{11.5\%\xspace}
\newcommand{\testDriftMthree}{9.1\%\xspace}

\newcommand{\koneUnusableGivenKtwoA}{11.4\%\xspace}
\newcommand{\ktwoNotAGivenKoneA}{6.5\%\xspace}
\newcommand{\koneSec}{1.6\xspace}
\newcommand{\ktwoSec}{19.7\xspace}
\newcommand{\ktwoSecRatio}{12\xspace}
\newcommand{\PdeadURL}{11.8\%\xspace}
\newcommand{\PdeadAbsentURL}{16.0\%\xspace}
\newcommand{\nHbestMiss}{14\xspace}
\newcommand{\nHbestMissShell}{9\xspace}
\newcommand{\nHbestMissSubst}{3\xspace}
\newcommand{\nMtwoFAOrdinary}{14\xspace}
\newcommand{\nMtwoFAShort}{5\xspace}
\newcommand{\nMtwoFAMoved}{3\xspace}
\newcommand{\nMtwoFACrossSite}{3\xspace}

\newcommand{\nGoldProbeUnusable}{31\xspace}
\newcommand{\nGoldPageUnusable}{68\xspace}

\newcommand{\recPageHtwo}{0.81\xspace}
\newcommand{\recPageHthree}{0.07\xspace}

\newcommand{\recPageHbest}{0.79\xspace}

\newcommand{\recProbeHthree}{0.97\xspace}
\newcommand{\primDiffPage}{+0.074\xspace}
\newcommand{\primDiffPageCI}{[0.016, 0.150]\xspace}
\newcommand{\autoPrecisionPage}{0.67\xspace}
\newcommand{\nTypeEPage}{11\xspace}
\newcommand{\nTypeEPageHbest}{8\xspace}
\newcommand{\nTypeEPageHthree}{1\xspace}
\newcommand{\nTypeEPageHfour}{4\xspace}
\newcommand{\nPilotPairs}{1,786\xspace}
\newcommand{\nPilotURLs}{1,606\xspace}
\newcommand{\nRobotsUnknown}{1,208\xspace}
\newcommand{\nRobotsUnknownOK}{289\xspace}
\newcommand{\nAutoTestUnusable}{272\xspace}
\newcommand{\autoDiffMtwoHbest}{0.05\xspace}
\newcommand{\moneMacroFone}{0.54\xspace}
\newcommand{\nMtwoFA}{26\xspace}
\newcommand{\nHbestFA}{11\xspace}
\newcommand{\faHtwoW}{0.047\xspace}
\newcommand{\polHbestTime}{5.8\xspace}
\newcommand{\polAlwaysTime}{20.6\xspace}

\newcommand{\polHbestUsable}{81.4\%\xspace}
\newcommand{\polAlwaysUsable}{80.3\%\xspace}

\newcommand{\polHbestRendered}{20\%\xspace}
\newcommand{\polHbestMissed}{7.8\%\xspace}
\newcommand{\polNeverMissed}{25.7\%\xspace}

\newcommand{\recHbestW}{0.82\xspace}
\newcommand{\faHbestW}{0.039\xspace}
\newcommand{\recMtwoW}{0.89\xspace}
\newcommand{\faMtwoW}{0.160\xspace}
\newcommand{\recMthreeW}{0.83\xspace}
\newcommand{\faMthreeW}{0.121\xspace}
\newcommand{\nRet}{3,307\xspace}
\newcommand{\retMedian}{0.985\xspace}
\newcommand{\retBelowNine}{17.3\%\xspace}
\newcommand{\retBelowHalf}{2.3\%\xspace}
\newcommand{\retFull}{43.9\%\xspace}
\newcommand{\nProbeTwoHundred}{172\xspace}
\newcommand{\probeAsContent}{28\xspace}
\newcommand{\probeAsError}{98\xspace}
\newcommand{\probeAsShell}{32\xspace}
\newcommand{\probeRedirectRoot}{42\xspace}
\newcommand{\probeSimHigh}{133\xspace}
\newcommand{\probeSimHave}{170\xspace}

\begin{document}
\setlength{\emergencystretch}{2em}

\title{When 200 OK Is Not an Answer: Silent Fetch Failures in Agentic Web Content Acquisition}

\author{Shuyang Zhang}
\affiliation{%
  \institution{The Hong Kong Polytechnic University}
  \city{Hong Kong}
  \country{China}}
\email{25049993g@connect.polyu.hk}

\begin{abstract}
Web agents and answer engines fetch pages at question time, and liveness checks of cited sources treat an HTTP 200 response as success. We measure how often that assumption fails and what it does to measurements built on fetched pages. We fetched \nItems{} URLs in \fetchDays{} days with a polite static client and a rendered browser client: pages from \nSitesS{} sites in five strata (llms.txt-linked pages, three Tranco rank bands and Markdown-serving sites), random non-existent paths on every site, and the \nPurls{} URLs cited by expert-verified claims in ExpertQA. We define a taxonomy of what a 200 response contains, label the static 200 responses automatically, and check the labels against \nGold{} items reviewed by an LLM-based assistant; a human check of \hcN{} of them retained every reviewed usability label after discussion. By the automatic labels, \unusableAllPct{} of the pages that a static client receives with status 200 do not contain the page's main content: app shells, bot challenges, block pages, error and placeholder pages, and partial content. On the held-out test split the reviewed sample lowers this estimate from \autoUnusableTestPct{} to \goldUnusablePct{}. \probeAllPct{} of the sites whose random-path probe was answered returned 200. The error propagates: for the cited sources, treating 200 as content overstates the share of claim--source pairs that can be checked against live content by \usableShiftPts{} points and inflates the apparent silent citation-drift rate from \driftAuto{} to \driftHzero{}. Cheap heuristics catch most of it. Text length, probe comparison and redirect-to-root, combined and tuned on held-out sites, recover \hbestGoldRecallPct{} of unusable reviewed responses (\recPageHbest{} on pages alone, without the random-path probes) at a false-alarm rate near 5\%. Content-reading detectors (a zero-shot LLM, a text classifier and a fine-tuned small language model) perform comparably: the best, the text classifier, gains five to seven points of recall at more than twice the false-alarm rate. We argue that agents and their evaluations should report fetch outcomes explicitly instead of inferring them from the status line, and will release the labelled data and code.
\end{abstract}


\maketitle

\section{Introduction}

Large language model (LLM) agents and answer engines increasingly read the Web live. A search-augmented assistant issues queries, fetches the top results, and writes an answer with citations~\cite{nakano2021webgpt,sayyid-ali2026sustainable}; evaluation pipelines check whether cited URLs are still live~\cite{rao2026detecting}, and attribution datasets pair claims with the evidence of the cited pages~\cite{malaviya2024expertqa}. Status-based liveness checks treat a fetch as a simple event: a response with status 200 is content, anything else is a failure~\cite{rao2026detecting}, and the same shortcut is easy to take whenever pages are fetched at scale. This paper asks how often that equivalence is wrong on today's Web, whether the wrong cases can be detected from the response itself, and how much the error moves the numbers that Web measurements and agent evaluations report. The Web relevance is direct: the populations are the pages agents are pointed at (llms.txt files, popular sites, cited sources), and the quantities affected (reachability, dead-link and drift rates) are standard Web measurements.

The status line is a weak signal for three reasons, each documented in isolation. Servers answer missing pages with a 200 ``soft 404'' or a redirect to the homepage~\cite{bar-yossef2004sic,prieto2013analysis}. Bot-management services often serve challenge and block pages, and refusals use status codes inconsistently~\cite{ansar2025web,gundelach2026detecting}. Many pages ship an application shell whose content arrives only after JavaScript runs, a long-standing problem for web archiving~\cite{brunelle2016impact,goel2022jawa}. A static fetch, the cheapest way to obtain a page, sees all three as a successful download. The recent measurement of answer engines' real-time fetching by Sayyid-Ali et al.~\cite{sayyid-ali2026sustainable} shows that engines fetch pages live at question time; it measures how much they fetch, not what the fetches return. Gundelach et al.~\cite{gundelach2026detecting} measure client-dependent blocking with explicit error codes and state that ``content degradation within HTTP 200 responses'' is outside their scope. That gap is the subject of this paper.

We make four contributions.
\begin{itemize}
\item \textbf{A fetch-outcome taxonomy for 200 responses} (Section~\ref{sec:taxonomy}) that separates main content from block pages, challenges, shells, error or substitute pages, parked pages and consent walls, together with a usability rule that also catches partial content by comparing a static fetch with a rendered one.
\item \textbf{A measurement} (Sections~\ref{sec:data}--\ref{sec:prevalence}) of \nItems{} fetches across six agent-relevant populations (\nEvalItems{} URLs fetched with both a static and a rendered client, plus \nRefProbes{} static-only reference probes), automatic labels for the static 200 responses, \nGold{} reviewed labels, a human check of \hcN{} of them and a browser check of the rendered reference. By the automatic labels a static 200 is unusable for \unusableAllPct{} of pages (the reviewed sample lowers the test-split estimate from \autoUnusableTestPct{} to \goldUnusablePct{}), with large differences between populations.
\item \textbf{A distortion analysis} (Section~\ref{sec:distortion}) on \nPairs{} claim--source pairs from ExpertQA: reading every 200 as content inflates the apparent silent-drift rate of cited sources from \driftAuto{} to \driftHzero{}, and status-based dead-link counts miss content that is gone behind a 200.
\item \textbf{A detector comparison} (Section~\ref{sec:detectors}) on site-disjoint splits, from the status code and five heuristics from crawling toolkits and the soft-404 literature to a zero-shot LLM, a text classifier and a fine-tuned small language model. A tuned combination of cheap heuristics recovers \hbestGoldRecallPct{} of unusable responses at a false-alarm rate near 5\%; the best content reader adds five to seven points of recall at more than twice the false-alarm rate, so cheap checks are the practical default for agent fetch tools.
\end{itemize}
All collection respected robots.txt (with one stated exception), rate limits and challenges; we never solved or bypassed a challenge (Section~\ref{sec:ethics}).

\section{Related Work}
\label{sec:related}

\textbf{Soft 404s and substitute pages.} Bar-Yossef et al.~\cite{bar-yossef2004sic} introduced probing a random sibling URL and comparing redirects and content to detect soft 404s, and found that about 29\% of dead links in their sample were soft. Prieto et al.~\cite{prieto2013analysis} reported that 7.35\% of servers answer unknown documents with 200 and built a content-based classifier for soft 404s. Our probe-comparison and redirect-to-root baselines follow this line. Parked domains, a substitute page that soft-404 probes cannot catch, have been measured and classified~\cite{vissers2015parking,alrwais2014understanding}.

\textbf{Blocking and bot walls.} Crawler choice changes what measurement studies see~\cite{ahmad2020apophanies,zeber2020representativeness,jueckstock2021towards}, and clients such as Tor face differential treatment~\cite{khattak2016you}. Ansar et al.~\cite{ansar2025web} detect refusal pages in Common Crawl from content with 147 regular expressions and show that status codes are used inconsistently. Gundelach et al.~\cite{gundelach2026detecting} find soft-block rates of 7--15\% on the Tranco top 10k depending on the browser configuration, counting only 403, 429 and 503 responses. Studies of robots.txt and AI crawlers explain why polite clients lose access~\cite{longpre2024consent,dinzinger2024longitudinal,kim2025scrapers,cui2025odyssey,liu2025somesite}. We add the responses that these status-based analyses count as successes, and we compare regex-style detection with learned content readers.

\textbf{Link rot and content drift.} Reference rot affects a large share of scholarly and legal citations~\cite{klein2014scholarly,zittrain2014perma}, and live content drifts from archived snapshots~\cite{jones2016scholarly}. Pew's large study counts a page as inaccessible only when it returns an error code, and calls this the most conservative estimate~\cite{chapekis2024online}. Rao et al.~\cite{rao2026detecting} check model-generated citation URLs with a status-based liveness rule. Ulloa et al.~\cite{ulloa2025beyond} show that scraping after the fact distorts measures of what users saw. Our distortion analysis quantifies the specific error introduced when 200 responses are taken as content.

\textbf{Agents and attribution.} Web agent benchmarks run on self-hosted sites~\cite{zhou2024webarena} or recorded snapshots of real sites~\cite{deng2023mind2web}, and attribution benchmarks score answers against evidence text assumed to be the cited page~\cite{gao2023enabling,li2024attributionbench,liu2023evaluating}. Live evaluations report captchas as a recurring failure~\cite{anupam2025browserarena}, work on answer-engine fetching studies its volume~\cite{sayyid-ali2026sustainable}, and benchmarks test the reliability of LLM-based search~\cite{wu2026paperask}. Crawling toolkits convert fetched pages to text or Markdown for agents~\cite{unclecode2024crawl4ai,jinaai2024reader,firecrawl2024firecrawl}; crawl4ai also ships a block detector, which we evaluate as a baseline.

\textbf{Rendering and main content.} JavaScript dependence has mainly been studied for web archiving~\cite{brunelle2016impact,goel2022jawa} and crawling of dynamic applications~\cite{mesbah2012crawling}, and boilerplate removal separates main content from navigation~\cite{kohlschutter2010boilerplate,barbaresi2021trafilatura}. Consent banners are measured as a privacy mechanism~\cite{rasaii2023thou}. We use these ideas to define shells, retention and consent walls as fetch outcomes.

\section{Fetch Outcomes}
\label{sec:taxonomy}

The unit of analysis is one fetch of one URL by an automated client. A fetch that ends without a response, with a robots.txt exclusion or with a status other than 200 is an \emph{explicit} outcome (dead: 404, 410 or no DNS record; blocked: 401, 403 or 429; failed: anything else). A 200 response is assigned one of seven types:
\begin{description}
\item[A, content:] the main content of the requested page;
\item[B, block:] an access-denied page;
\item[C, challenge:] a JavaScript or bot check or a captcha;
\item[D, shell:] navigation, skeleton or application shell without the main content;
\item[E, error or substitute:] an error or not-found page, or a different page such as the homepage;
\item[F, parked:] a parked domain or placeholder;
\item[G, consent wall:] only a cookie or consent wall.
\end{description}
Table~\ref{tab:types} gives examples and counts. Types B--G are \emph{unusable}: an agent that reads them has not read the page. The types are deliberately coarse. They are defined by what a reader of the response can use, not by the server's intent, so a geographic block and a bot block are both type B, and a login page served instead of a deep link is type E unless the rendered client receives the same page.

\begin{table}[t]
\caption{Outcome types for 200 responses, with examples from the reviewed sample. Auto: automatic type of all 4,560 static 200 responses (pages and probes). Rev.: reviewed type of the 300 sampled items.}
\label{tab:types}
\small\setlength{\tabcolsep}{3pt}
\begin{tabular}{@{}llp{3.55cm}rr@{}}
\toprule
& Type & Example & Auto & Rev.\\
\midrule
A & Content & article, product, documentation or listing text & 3,575 & 199 \\
B & Block & geo-block (``location not supported''); WAF ``request unsuccessful'' & 83 & 3 \\
C & Challenge & Cloudflare Turnstile; publisher ``Client Challenge''; image captcha & 159 & 12 \\
D & Shell & product grid that only says ``Loading...''; live scores missing; empty body & 354 & 48 \\
E & Error or substitute & random path redirected to the homepage; unrelated page served at a random path & 341 & 31 \\
F & Parked & hosting placeholder; parked domain redirecting to a lander & 27 & 6 \\
G & Consent wall & publisher privacy gate in front of the article & 21 & 1 \\
\bottomrule
\end{tabular}
\end{table}
 Type A can still be incomplete when content depends on scripts. We therefore pair every static fetch with a rendered fetch of the same URL and compute the \emph{retention} $r$: the shares of the rendered page's sentences and of its numeric facts that appear in the static text, averaged over the unit kinds present, after removing consent-banner text. A static 200 response is unusable if (i) its type is B--G, (ii) both responses are type A and $r<0.5$, or (iii) it answers a random non-existent path (a soft 404). We report sensitivity at $r<0.3$ and $r<0.7$. When the rendered fetch is itself not type A, no reference exists and a type-A static response counts as usable.

\section{Data Collection}
\label{sec:data}

\begin{table}[t]
\caption{Populations. Pages are fetched with both clients; every site in S1--S5 also receives two random-path probes.}
\label{tab:pops}
\small
\begin{tabular}{@{}llrr@{}}
\toprule
ID & Population & Sites & Pages \\
\midrule
S1 & Links listed in llms.txt files~\cite{gerrish2026llmstxt} & \sitesSone & \pagesSone \\
S2 & Tranco~\cite{le-pochat2019tranco} top 10k & \sitesStwo & \pagesStwo \\
S3 & Markdown-serving sites in the top 10k & \sitesSthree & \pagesSthree \\
S4 & Tranco ranks 10,001--100,000 & \sitesSfour & \pagesSfour \\
S5 & Tranco ranks 100,001--1,000,000 & \sitesSfive & \pagesSfive \\
P & URLs cited in ExpertQA~\cite{malaviya2024expertqa} & \sitesP & \pagesP \\
\bottomrule
\end{tabular}
\end{table}

\textbf{Populations.} Table~\ref{tab:pops} lists the six populations; all sampling used one fixed seed. S1 draws two links from each sampled site's llms.txt file~\cite{howard2024llmstxt}, using a public corpus of llms.txt files~\cite{gerrish2026llmstxt} and only sites with at least two links that the corpus recorded as answering 200, so S1 is biased towards working links. S2, S4 and S5 sample sites from three Tranco rank bands; for each we fetch the homepage and one random same-site link from the rendered homepage. S3 contains the sites in the Tranco top 10k that answered a request with \texttt{Accept: text/markdown} with Markdown (\sitesSthree{} sites, after excluding pilot sites). P contains the cited URLs behind ExpertQA claims whose support was judged complete or partial and whose evidence passage has at least 100 characters (\nPairs{} claim--URL pairs, at most three per URL, \nPurls{} URLs). Sites used in our pilot were excluded and sites that did not resolve at sampling time were replaced. The cited sources overlap with an earlier pilot of ours on ExpertQA (\nPilotPairs{} of the pairs, \nPilotURLs{} of the URLs), whose drift-judge prompt we reuse; all URLs were fetched afresh. Every site in S1--S5 also receives two random paths of 16 hexadecimal characters: one is an item (a soft-404 test), the other is fetched statically only, is not labelled, and serves as the reference for probe comparison.

\textbf{Clients.} The static client (K1) is a single GET with redirects followed and a 20-second timeout. The rendered client (K2) loads the page in headless Chromium through Playwright~\cite{microsoft2019playwright}, waits for network idle (up to 8 seconds), and saves the visible text and a screenshot. Collection ran from \fetchStart{} to \fetchEnd{} from one vantage point. robots.txt was honoured following RFC~9309~\cite{koster2022rfc9309}, with one exception: when the robots.txt request itself got no answer (no DNS record, connection failure or timeout), RFC~9309 says to assume a full disallow, but we still attempted the page, so that dead and unreachable hosts remain countable (\nRobotsUnknown{} items, of which \nRobotsUnknownOK{} received a static 200). Each host saw at most one request per second, and contexts were fresh for every item, so no cookies carried across sites.

\textbf{Labels.} Every K1 response with status 200, and the K2 response of the same item when it returned text, was labelled with the seven types by an LLM (deepseek-v4-flash, temperature 0) from the URL, final URL, title and the first 3,000 characters of visible text; for K1 the prompt also showed an excerpt of K2's text as context. The prompt was fixed before any label was seen. Retention uses sentence and numeric-fact units.

\textbf{Splits and statistics.} Items are assigned to train (60\%), validation (20\%) or test (20\%) by a seeded hash of their registrable domain, so all pages and probes of a site, and all cited URLs on one domain, share a split. Confidence intervals resample registrable domains (or sites) with replacement, 2,000 times.

\textbf{Reviewed labels and browser check.} We drew \nGold{} items from the test split (Section~\ref{sec:detectors}), stratified by population and automatic usability. Each was reviewed blind to its automatic label, from the URL, status, final URL and short excerpts of K1's and K2's visible text (usually 500 characters, longer where needed), and given a type and a usability label; \nGoldUnsure{} items could not be decided and are excluded from usability metrics. The sample has twelve cells (six populations by automatic usability) of 25 items each; cells with fewer items contribute all of them and the remainder is shared among the others, and every item carries its cell's sampling weight so that estimates can be reweighted to the whole test split. This review was carried out by an LLM-based assistant (Claude), a different model family from the labeller and the zero-shot detector, under written rules (Appendix~\ref{app:review}). The LLM review is text-based: the K2 screenshot was consulted for one item only. \emph{Human check.} An author labelled \hcN{} reviewed test pages (30 reviewed-unusable, 30 reviewed-usable, drawn with a fixed seed under a plan written before labelling), blind to both labels, from K1's visible text with K2's text and screenshot as reference. In this blind pass, usability agreed with the reviewed label for \hcAgreeN{} of \hcNscored{} decidable items (\hcAgreePct{}, \hcAgreeWPct{} reweighted to the \hcFrameN{} reviewed test pages; Cohen's $\kappa=\hcKappa{}$; type $\kappa=\hcTypeKappa{}$). All \hcDisagree{} usability disagreements were then resolved by discussion between the author and the reviewer, every one in favour of the reviewed label: \hcCauseRendered{} came from judging the rendered page or the live site instead of the static text, \hcCauseRedirect{} from missing a redirect to an unrelated site, \hcCauseForeign{} from an error page in another language and \hcCausePlaceholder{} from a hosting placeholder. The final type differed from the reviewed type for \hcFinalTypeDiff{} items, all unusable either way. As the reviewer took part in the discussion, the blind pass is the independent comparison. Of the unusable reviewed items, \nGoldProbeUnusable{} are random-path probes, which are unusable by rule (iii); we therefore also report detector recall on the \nGoldPageUnusable{} unusable pages alone. To test the rendered reference (not the reviewed labels), an author opened \nBrowser{} random reviewed items in a normal desktop browser and judged whether K2's text contained the main content seen in the browser. K2 missed content that the browser showed for \browserMiss{} of \browserN{} pages (\browserMissPct{}); the rule for scoring the author's notes was fixed after the answers were read.

\section{How Often 200 Is Not Content}
\label{sec:prevalence}

\begin{figure}[t]
\centering
\includegraphics[width=\columnwidth]{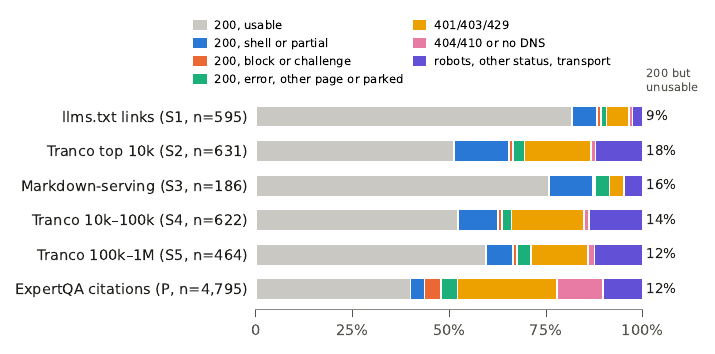}
\caption{Outcome of the static fetch for every page, by population. Grey: status 200 with main content. Blue, orange and green: status 200 without usable content (automatic labels, $r<0.5$); their total share of all pages is printed on the right. Other colours: explicit outcomes.}
\label{fig:outcomes}
\end{figure}

\begin{table}[t]
\caption{Static outcomes by population, in \%. 200: share of pages answered with 200. Unusable: share of those 200 responses that are unusable (automatic labels, $r<0.5$), with a 95\% site-bootstrap interval, and reweighted from the reviewed test sample (test split only, 36--57 pages per population, no interval). Probe: share of sites whose random-path probe was answered that answered with 200. Blocked: pages answered 401/403/429. Agree: usability agreement of automatic and reviewed labels on the reviewed items (pages and probes).}
\label{tab:prev}
\small\setlength{\tabcolsep}{2.6pt}
\begin{tabular}{@{}lrrrrrrrr@{}}
\toprule
 & Pages & 200 & Unus. & CI & Reviewed & Probe & Blocked & Agree\\
\midrule
S1 & 595 & 90.6 & 9.8 & 7--13 & 5.4 & 10.5 & 5.9 & 96 \\
S2 & 631 & 69.4 & 26.3 & 22--31 & 23.1 & 13.3 & 17.1 & 79 \\
S3 & 186 & 91.4 & 17.1 & 11--24 & 7.0 & 17.3 & 3.8 & 87 \\
S4 & 622 & 66.1 & 20.9 & 17--25 & 19.0 & 13.7 & 18.6 & 88 \\
S5 & 464 & 71.1 & 16.4 & 12--21 & 12.5 & 13.1 & 14.7 & 89 \\
P & 4,795 & 52.1 & 23.2 & 19--28 & 25.3 & -- & 25.7 & 91 \\
\bottomrule
\end{tabular}
\end{table}

Figure~\ref{fig:outcomes} and Table~\ref{tab:prev} show the static outcome of all \nPages{} pages. \kOneTwoHundredPct{} return 200. Of these, \unusableAllPct{} are unusable by the automatic labels, from \unusableSonePct{} for llms.txt links to \unusableStwoPct{} for the Tranco top 10k (\unusablePPct{} for cited sources). The reviewed sample confirms the pattern but lowers the level: reweighted to the test split, \goldUnusablePct{} (95\% CI \goldUnusableCI) of static 200 page responses are unusable, against \autoUnusableTestPct{} by the automatic labels on the same split.\footnote{The automatic share is higher on the test split (\autoTestPagesPct{} of its static 200 pages) than on all pages (\unusableAllPct{}) because the split is by registrable domain and a few large, mostly unusable sites fell into it, notably all \nTestMdpi{} mdpi.com pages, \nTestMdpiUnusable{} of them unusable; without mdpi.com the test-split share is \autoTestNoMdpiPct{}.} Under both label sources the same three populations are worst (the Tranco top 10k, the 10k--100k band and the cited sources) and llms.txt links are best, partly because S1 was restricted to links recorded as working. The automatic labels find nearly all unusable responses (recall \autoRecall{}) but also flag many usable ones (precision \autoPrecision{}, and \autoPrecisionPage{} on pages alone); their type labels agree with the reviewed types on \autoTypeAcc{} of items (Cohen's $\kappa$ = \autoTypeKappa{}).

\textbf{What the failures are.} Shells and partial content (type D, consent walls and low retention) are the largest group in every Tranco band: an application frame with navigation and footer but without the article, product or listing. Challenges and block pages concentrate in the cited sources, where \PaccessPct{} of 200 responses are challenge or block pages; four document and scholarly hosts (\PaccessTopFour) account for \PaccessTopFourPct{} of them, so the rate depends heavily on which publishers a pipeline cites. Error and substitute pages (type E) include homepages served for deep links and generic ``page not found'' templates with status 200. The threshold matters little: at $r<0.3$ and $r<0.7$ the unusable share of static 200 pages is \unusableLoPct{} and \unusableHiPct{}.

\textbf{Soft 404s.} \probeAllPct{} of the sites that answered the random-path probe returned 200 (S1 \probeSonePct{}, S2 \probeStwoPct{}, S3 \probeSthreePct{}, S4 \probeSfourPct{}, S5 \probeSfivePct{}). Every such response is a soft 404 by construction, but not every one says so. In an exploratory breakdown of the \nProbeTwoHundred{} soft-404 responses, the labeller called \probeAsError{} an error or substitute page, \probeAsShell{} a shell, and \probeAsContent{} content: a homepage or listing served at a random path reads like a normal page. Only \probeRedirectRoot{} redirected to the root, while \probeSimHigh{} of \probeSimHave{} had word 3-gram similarity of at least 0.8 with the site's second probe. A soft 404 is thus partly a property of the request--response pair rather than of the response alone.

\begin{figure}[t]
\centering
\includegraphics[width=\columnwidth]{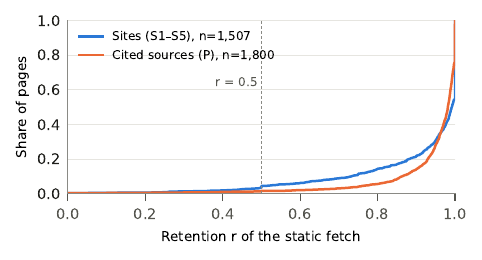}
\caption{Cumulative distribution of retention $r$ (mean share of the rendered page's sentences and numeric facts found in the static text) for pages where both clients returned content.}
\label{fig:retention}
\end{figure}

\textbf{Partial content (exploratory).} Where both clients returned content, static fetches usually keep nearly all of it (Figure~\ref{fig:retention}): over \nRet{} such pages the median retention is \retMedian{}, and \retFull{} keep at least 99\% of the rendered units. The tail is long, though: \retBelowNine{} of pages lose more than a tenth of their sentences and facts, while only \retBelowHalf{} fall below the $r<0.5$ cut-off that our usability rule uses. Partial loss is therefore common but usually moderate, and complete shells (type D) are the main form of script dependence.

\textbf{Explicit failures are common too.} \blockedAllPct{} of all pages answered the static client with 401, 403 or 429, and \blockedPPct{} of cited sources. For \kTwoRescuePct{} of these blocked pages, the rendered client received status 200; a 200 is not yet content, and these responses are exposed to the 200-but-unusable failures above.

\textbf{Static versus rendered (exploratory).} The analyses from here on were added after the main results (Appendix~\ref{app:deviations}). Rendering is an obvious remedy, so we compare the two clients on the same URLs. When the rendered page has main content (K2 type A), the static response is still unusable for \koneUnusableGivenKtwoA{} of pages, mostly shells, challenges served only to the static client, and low retention; conversely, for \ktwoNotAGivenKoneA{} of pages whose static response is content, the rendered fetch is not, mostly because rendering failed or returned no text, or because the headless browser received a block page, a shell or a consent overlay. Rendering is also expensive: the median rendered fetch took \ktwoSec{}~s against \koneSec{}~s for the static one, roughly \ktwoSecRatio{} times longer, so a pipeline that fetches statically first needs to know when to escalate.

\section{Detecting Unusable Responses}
\label{sec:detectors}

\textbf{Setup.} Detectors judge the static response only, as an agent would before deciding to render, retry or skip a page. Detectors use the site-disjoint splits of Section~\ref{sec:data}. Thresholds are chosen on validation as the setting with the highest recall on unusable items at a false-alarm rate of at most 5\% on usable items (automatic labels). We report recall and false-alarm rate on the reviewed test sample, with domain-bootstrap 95\% confidence intervals.

\textbf{Detectors.} H0 trusts the status code. H1 is the \texttt{is\_blocked} check of crawl4ai~\cite{unclecode2024crawl4ai}. H2 flags visible text shorter than a threshold. H3 compares the page with the same site's second random-path probe (word 3-gram Jaccard similarity) for non-root URLs~\cite{bar-yossef2004sic}; it applies only when that probe was answered with 200, so it can fire only on sites that soft-404, and the standalone threshold tuned on validation is a similarity of 0, i.e.\ any non-root URL of such a site. H4 flags a redirect from a deep URL to the root. H5 matches challenge and block strings~\cite{ansar2025web}. H-best is the OR of H1--H5 with all thresholds and switches tuned jointly on validation; the selected configuration is redirect-to-root, or probe similarity of at least 0.3, or fewer than 930 characters of text (H1 and H5 switched off). Among content readers, M1 is a zero-shot LLM (Qwen3-32B~\cite{yang2025qwen3}, thinking off, a different family from the labeller) asked for the type with the labelling prompt minus the rendered context; any type other than A counts as unusable. We also report its type agreement with the reviewed labels (macro F1 \moneMacroFone{}). M2 is logistic regression~\cite{pedregosa2011scikit} on TF-IDF word 1--2-grams and character 3--5-grams of title, text and URL plus the heuristic features. M3 fine-tunes a Qwen3 sequence classifier on the first 1,024 tokens of title, text and URL (the URL is appended after the text, so it is cut off for long pages; two epochs, learning rate $2\times10^{-5}$). As fixed in the protocol, Qwen3-0.6B was trained first and, because it fell short of M2 on validation, Qwen3-1.7B was trained once; the 1.7B model was better on validation and is M3.

\begin{table}[t]
\caption{Detectors on the test split. Recall on unusable and false alarm (FA) on usable items, on the reviewed sample (99 unusable, 195 usable) and on all test items with automatic labels (272 / 684). Thresholds fixed on validation at FA $\le$ 5\%. CI: 95\% domain bootstrap. Last column: recall at FA $\le$ 5\% with the threshold chosen on the reviewed sample itself (descriptive only). Pages: recall on the 68 unusable pages, excluding the 31 random-path probes (FA is unchanged, as all usable items are pages).}
\label{tab:detectors}
\small\setlength{\tabcolsep}{3pt}
\begin{tabular}{@{}lrrrrrrr@{}}
\toprule
& \multicolumn{4}{c}{Reviewed sample} & \multicolumn{2}{c}{Automatic} & Oracle\\
\cmidrule(lr){2-5}\cmidrule(lr){6-7}
Detector & Recall & CI & Pages & FA & Recall & FA & Recall\\
\midrule
H0 & 0.00 & 0.00--0.00 & 0.00 & 0.00 & 0.00 & 0.00 & -- \\
\midrule
H1 & 0.25 & 0.14--0.37 & 0.25 & 0.02 & 0.26 & 0.03 & -- \\
H2 & 0.76 & 0.66--0.85 & 0.81 & 0.07 & 0.65 & 0.04 & 0.66 \\
H3 & 0.35 & 0.26--0.46 & 0.07 & 0.03 & 0.15 & 0.02 & 0.35 \\
H4 & 0.10 & 0.05--0.16 & 0.06 & 0.00 & 0.08 & 0.00 & -- \\
H5 & 0.06 & 0.00--0.16 & 0.09 & 0.02 & 0.11 & 0.01 & -- \\
H-best & 0.86 & 0.78--0.93 & 0.79 & 0.06 & 0.74 & 0.02 & 0.84 \\
\midrule
M1 & 0.57 & 0.45--0.68 & 0.50 & 0.06 & 0.48 & 0.03 & -- \\
M2 & 0.91 & 0.84--0.97 & 0.87 & 0.13 & 0.78 & 0.05 & 0.69 \\
M3 & 0.82 & 0.73--0.90 & 0.82 & 0.13 & 0.76 & 0.06 & 0.76 \\
\bottomrule
\end{tabular}
\end{table}

\begin{figure}[t]
\centering
\includegraphics[width=\columnwidth]{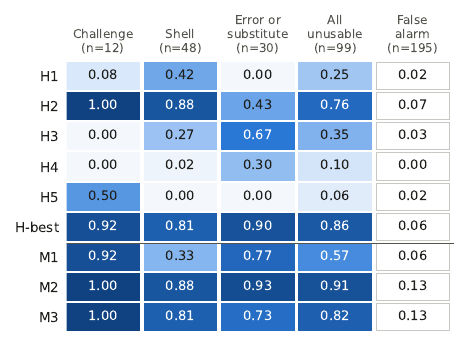}
\caption{Recall by reviewed type (types with at least 12 unusable items) and false-alarm rate on usable items, reviewed test sample. Thresholds fixed on validation.}
\label{fig:types}
\end{figure}

\textbf{Results.} Table~\ref{tab:detectors} reports all detectors on the \goldNun{} unusable and \goldNus{} usable reviewed items. Status codes detect nothing by definition. Text length is the strongest single signal (\recHtwo{}; \recPageHtwo{} on pages) because shells, challenges and block pages are short. The crawl4ai block detector finds \recHone{} of unusable responses, mostly shells, and challenge strings find half of the challenges but little else. Probe comparison and redirect-to-root mainly find the random-path probes themselves: H3 catches \recProbeHthree{} of the probes but only \recPageHthree{} of unusable pages, and of the \nTypeEPage{} substitute or error pages in the sample, H3 catches \nTypeEPageHthree{} and H4 \nTypeEPageHfour{}. Their tuned combination, H-best, reaches a recall of \recHbest{} at a false-alarm rate of \faHbest{} (\recPageHbest{} on pages), and recovers most shells (\nHbesttypeD{} of \goldTypeD{}), challenges (\nHbesttypeC{} of \goldTypeC{}) and substitute or error items (\nHbesttypeE{} of \goldTypeE{}; \nTypeEPageHbest{} of the \nTypeEPage{} that are pages).

Content readers do not change this picture. The zero-shot LLM (M1) is conservative: its type labels match the reviewed types on \moneTypeAcc{} of items, but it calls most shells content (recall \recMonetypeD{} on type D), so it finds only \recMone{} of unusable responses. The text classifier (M2) has the highest recall, \recMtwo{}, but its false-alarm rate is \faMtwo{}, more than twice H-best's, and \faMtwoP{} on usable cited sources; the fine-tuned Qwen3-1.7B (M3) behaves similarly (\recMthree{} at \faMthree{}). M2's recall exceeds H-best's by \primDiff{} (95\% CI \primDiffCI{}; \primDiffPage{} on pages alone, \primDiffPageCI{}), and by \autoDiffMtwoHbest{} on the larger automatically labelled test set (\nAutoTestUnusable{} unusable items; M2 \autoRecMtwo{} at a false-alarm rate of \autoFaMtwo{}, H-best \autoRecHbest{} at \autoFaHbest{}); the protocol's primary comparison had set a margin of 0.10 for this gain. With thresholds set on the reviewed sample itself to equalise the false-alarm rate at 5\%, H-best (\oracleHbest{}) recalls more than M2 (\oracleMtwo{}) and M3 (\oracleMthree{}) in this descriptive comparison. Figure~\ref{fig:types} breaks recall down by type; the error-or-substitute column is dominated by probes. Reweighting the reviewed items to the composition of the test split gives the same picture (H-best \recHbestW{} at \faHbestW{}, M2 \recMtwoW{} at \faMtwoW{}, M3 \recMthreeW{} at \faMthreeW{}).

\textbf{Error analysis (exploratory).} We coded every reviewed error of the two strongest detectors into exclusive categories. H-best misses \nHbestMiss{} unusable responses: \nHbestMissShell{} are shells that carry enough navigation, SEO or licence text to pass the length cut-off (``Loading...'' product grids, sports pages whose scores load by script), \nHbestMissSubst{} are substitute or error pages among the cited sources, which do not redirect to the root and have no probe to compare with (a deep link redirected to an agency homepage or to an image), one is a challenge and one is partial content. M2 raises \nMtwoFA{} false alarms on usable pages (H-best: \nHbestFA{}). \nMtwoFAOrdinary{} are ordinary articles, listings and homepages with no visible defect; \nMtwoFAShort{} are short or unusual but complete pages (a directory index, a one-line response, a plain-text file); \nMtwoFAMoved{} are articles reached through a redirect to a moved URL; \nMtwoFACrossSite{} redirect to another site's page that the rendered client also receives; and one is a URL whose own content is a location-not-supported notice. Deciding the moved and cross-site cases requires comparing the response with what the requested URL should return, which neither detector does.

On the reviewed sample, H2, H-best, M2 and M3 all exceed the 5\% false-alarm rate set on validation with automatic labels, consistent with those labels over-flagging unusability. The stratified sample adds to this, since it over-represents items that the automatic labels call unusable: reweighted, the false-alarm rates of H2 and H-best are \faHtwoW{} and \faHbestW{}.

\textbf{When to render (exploratory).} The practical use of a detector is to decide when a cheap static fetch must be escalated to a rendered one. Table~\ref{tab:policy} simulates this on the test-split pages with the automatic labels. Rendering only the pages H-best flags (\polHbestRendered{} of them) yields as much usable content as rendering everything (\polHbestUsable{} against \polAlwaysUsable{}, because some pages are usable only statically) at \polHbestTime{}~s per page instead of \polAlwaysTime{}~s, and reduces the unusable responses passed on silently from \polNeverMissed{} to \polHbestMissed{}. Learned detectors trade a few more renders for a slightly lower miss rate.

\begin{table}[t]
\caption{Fetch policies on the 920 test-split pages with a static 200 (automatic labels as outcome). Usable: page content obtained. Missed: unusable static response used without escalation. Rendered: share of pages rendered. Time: mean seconds per page (static plus rendered when escalated).}
\label{tab:policy}
\small\setlength{\tabcolsep}{3pt}
\begin{tabular}{@{}lrrrr@{}}
\toprule
Policy & Usable \% & Missed \% & Rendered \% & Time (s)\\
\midrule
Never render & 74.3 & 25.7 & 0.0 & 2.5 \\
Always render & 80.3 & 0.0 & 100.0 & 20.6 \\
Render if H-best flags & 81.4 & 7.8 & 19.6 & 5.8 \\
Render if M2 flags & 81.6 & 6.4 & 22.9 & 6.6 \\
Render if M3 flags & 81.8 & 6.2 & 23.6 & 6.7 \\
\bottomrule
\end{tabular}
\end{table}

\section{Distortion of Downstream Measurements}
\label{sec:distortion}

\begin{table}[t]
\caption{Cited sources (pairs). Share of pairs whose static fetch is live and usable or live and unusable, and silent-drift share among usable live pairs whose 2023 passage supported the claim, in \%, under each way of judging 200 responses. M2 and M3 were trained on train-split items, so they are shown on the test split only. Dead: 11.0\% of all pairs and 12.0\% of test pairs; failed (blocked, other errors, transport failures, robots.txt exclusions): 36.4\% and 31.8\%.}
\label{tab:distortion}
\small\setlength{\tabcolsep}{3pt}
\begin{tabular}{@{}llrrrrrr@{}}
\toprule
& & Status only & H-best & M1 & M2 & M3 & Labels\\
\midrule
All cited & Live, usable & 52.6 & 45.1 & 45.4 & -- & -- & 40.5 \\
 & Live, unusable & -- & 7.5 & 7.2 & -- & -- & 12.2 \\
 & Silent drift & 21.9 & 10.2 & 11.4 & -- & -- & 5.8 \\
\midrule
Test split & Live, usable & 56.1 & 42.7 & 46.8 & 41.0 & 40.4 & 38.3 \\
 & Live, unusable & -- & 13.5 & 9.3 & 15.1 & 15.7 & 17.8 \\
 & Silent drift & 31.9 & 12.6 & 18.1 & 11.5 & 9.1 & 6.3 \\
\bottomrule
\end{tabular}
\end{table}

ExpertQA lets us measure the effect of fetch-outcome errors on a concrete pipeline: re-fetching cited sources to check whether they are live~\cite{rao2026detecting} and still support the claim they were cited for~\cite{malaviya2024expertqa}. For each of the \nPairs{} pairs we take the static fetch of the cited URL. A judge model (the labelling model with a fixed prompt) compares the claim with the 2023 evidence passage and with the best-matching 4,000-character window of the current page. A pair \emph{silently drifts} if the passage supported the claim but the live page, fetched with status 200 and judged usable, no longer does. Table~\ref{tab:distortion} recomputes these quantities under each way of deciding whether a 200 response is content. \deadP{} of pairs point to dead URLs and \failedP{} fail explicitly (blocked, other errors, transport failures or robots.txt exclusions); these are the same under every scheme. The difference lies in the 200 responses. Counting every 200 as content, \liveUsableHzero{} of pairs can be checked against a live page; with outcome labels, \liveUsableAuto{}. The effect on drift is larger. Of the live pairs counted as usable whose 2023 passage supported the claim (\driftBaseHzero{} pairs under the status-only view, \driftBaseAuto{} with labels), \driftHzero{} appear to have silently drifted when every 200 is read as the page, against \driftAuto{} when unusable responses are set aside: \driftRatio{} times as much apparent drift. Of the \nDriftHzero{} apparent drifts under the status-only view, \nDriftUnusable{} (\driftUnusableShare{}) are responses labelled unusable: \nDriftAbsent{} error, substitute or parked pages, \nDriftAccess{} block or challenge pages and \nDriftDegraded{} shells or partial pages. Heuristic and learned detectors also remove much of this inflation (H-best \driftHbest{}, M1 \driftMone{}; on the test split, M2 \testDriftMtwo{} and M3 \testDriftMthree{}).

\textbf{By answer system (exploratory).} The inflation differs by how the answers were produced (Table~\ref{tab:systems}). ExpertQA's systems retrieve sources either from Google search or from Sphere, a static web corpus, and either before writing (retrieve-and-read) or after (post hoc)~\cite{malaviya2024expertqa}. Status-only drift is about three times the label-based value for three systems and five times for retrieve-and-read over Sphere, although the label-based values rest on only 6 to 26 drift events per system. Sphere-sourced citations are also live and usable less often, so a comparison of answer systems by re-fetched support would rank them partly by the fetchability of their sources.

\begin{table}[t]
\caption{Cited sources by ExpertQA answer system. Live: static 200, in \% of pairs; usable: also labelled usable. Drift: silent-drift share counting every 200 as content (status) and only usable responses (labels).}
\label{tab:systems}
\small\setlength{\tabcolsep}{3pt}
\begin{tabular}{@{}lrrrrr@{}}
\toprule
 & & \multicolumn{2}{c}{Pairs \%} & \multicolumn{2}{c}{Drift \%}\\
\cmidrule(lr){3-4}\cmidrule(lr){5-6}
System & Pairs & Live & Usable & Status & Labels\\
\midrule
Post hoc, Google & 1,888 & 55.5 & 45.3 & 11.5 & 3.8 \\
Retrieve-and-read, Google & 1,553 & 57.2 & 46.9 & 16.5 & 5.3 \\
Post hoc, Sphere & 1,517 & 47.7 & 33.4 & 30.1 & 10.3 \\
Retrieve-and-read, Sphere & 1,198 & 48.4 & 33.4 & 33.3 & 6.2 \\
\bottomrule
\end{tabular}
\end{table}

The same error affects dead-link measurement (exploratory). Counted by status, \PdeadURL{} of the cited URLs are dead; counting 200 responses automatically labelled as error, substitute or parked pages as well, \PdeadAbsentURL{} no longer serve the cited content. A status-based link-rot study, such as the conservative design of Pew~\cite{chapekis2024online}, would miss the difference.

\section{Discussion}
\label{sec:discussion}

\textbf{Report fetch outcomes, not status codes.} Every number in Sections~\ref{sec:prevalence} and~\ref{sec:distortion} that depends on ``the page was fetched'' moves when 200 is replaced by an outcome label. We suggest three practices. (1) Agent fetch tools should return an outcome type (content, block, challenge, shell, substitute, parked, consent) with every response and escalate to rendering on a flag (Table~\ref{tab:policy}); a length cut-off with probe and redirect checks is enough to start. (2) Evaluations that re-fetch cited sources should publish the outcome distribution of their fetches, judge support only on usable responses, and report how much of their headline rate rests on 200 responses that were not inspected. (3) Link-rot and liveness studies should report content-absent rates (dead plus substitute and parked 200s) next to status-based dead rates.

\textbf{Rendering is not a cure.} Rendering turned \kTwoRescuePct{} of explicit static blocks into 200 responses and gave main content for \shellRescuePct{} of static shells, but the rendered reference is itself not type A for \kTwoNotAPct{} of the static-200 items, and a browser check found that it missed content in \browserMissPct{} of pages. Outcome labelling remains necessary after rendering.

\textbf{Limitations.} All fetches came from one vantage point, and block and challenge rates depend on the client, its IP address and time; the absolute share of each type is therefore specific to this setup, and a residential or cloud vantage point or a different week would shift it, although we expect the same types to recur. The drift judge is an LLM applied under every scheme; because each scheme passes a different subset of pages to it, its errors need not cancel between schemes, so the gap between status-only and label-based drift rates may include judge error as well as the effect of unusable pages.

\section{Ethics}
\label{sec:ethics}

Both clients sent the same desktop Chrome user-agent string and therefore did not announce themselves as crawlers; they obeyed the robots.txt rules that apply to all user agents (RFC~9309, except that hosts whose robots.txt could not be reached were still attempted once), waited at least one second between requests to a host, did not log in, and never attempted to solve or bypass a challenge; challenge pages are recorded as outcomes. The load was a few requests per site for S1--S5; for cited sources at most 161 URLs fell on one host (en.wikipedia.org), fetched at the same one-second spacing. We will release URLs, labels, derived features and code; raw page bodies are kept for audit and are not redistributed.

\section{Conclusion}

An HTTP 200 is not an answer. Across six populations of pages that agents fetch, a sizeable share of static 200 responses carry shells, challenges, block pages, substitute pages or partial content, and a further share of sites answer a random non-existent path with 200. Treating these as content inflates liveness estimates and distorts drift estimates. Cheap heuristics catch much of it, and learned content readers perform comparably at a higher false-alarm rate. Outcome-aware fetching and reporting is a small change with a measurable effect on what agents read and on what their evaluations conclude.

\bibliographystyle{ACM-Reference-Format}
\bibliography{refs}

\appendix
\section{Prompts}
\label{app:prompts}

\textbf{Type labeller} (deepseek-v4-flash, temperature 0; M1 uses the same text without the rendered excerpt). Placeholders are in braces.
\begin{footnotesize}
\begin{verbatim}
You are labelling what an automated web client received
when it fetched a URL.

Decide which ONE type describes the fetched content:
A = the main content of the requested page (article, product,
    documentation, listing, homepage content, etc.)
B = blocked or access denied (e.g. "Access Denied",
    "403 Forbidden", "you have been blocked")
C = a challenge: JavaScript or bot check, captcha, "checking
    your browser", "verify you are human"
D = navigation, skeleton or app shell only, without the page's
    main content (e.g. menus and footer only, "loading...",
    "enable JavaScript")
E = an error, not-found, or a different page than requested
    (homepage or generic page instead of the requested one)
F = a parked domain or placeholder ("domain for sale",
    "coming soon", default server page)
G = only a cookie or consent wall, without the main content

URL requested: {url}
Final URL: {final_url}
HTTP status: {status}
Title: {title}

Visible text (first 3,000 characters):
<<<
{text}
>>>
[static responses only:] For context, an excerpt of the same
URL as rendered by a full browser (first 1,500 characters):
<<<
{k2}
>>>
Return only JSON:
{"type": "A" | "B" | "C" | "D" | "E" | "F" | "G"}
\end{verbatim}
\end{footnotesize}

\textbf{Drift judge} (same model). \texttt{TEXT} is either the 2023 evidence passage or the best-matching 4,000-character window of the current page, chosen by token overlap with the claim and passage.
\begin{footnotesize}
\begin{verbatim}
Does the TEXT support the CLAIM? Answer yes if the text states
the claim's content, partial if it supports only part of it,
no if it does not support it.

CLAIM: {claim}

TEXT:
{text}

Return only JSON:
{"support": "yes" | "partial" | "no"}
\end{verbatim}
\end{footnotesize}

\section{Reviewing Rules}
\label{app:review}

The reviewed labels follow written conventions. They were assigned from the URL, status, final URL and visible-text excerpts of both clients, blind to the automatic labels. A redirect to a page that the rendered client also receives for the requested URL is content (A). A random path or deep link that lands on the homepage or a generic page is a substitute (E). A page whose text is only a title, navigation or a script loader is a shell (D). Bot checks, captchas and silent script challenges are challenges (C); geographic blocks are blocks (B); hosting placeholders and parked domains are parked (F). Character-set artefacts that appear only in the review viewer are not fetch defects. Items whose usability could not be decided from the text were marked unsure and excluded from usability metrics.

\section{Protocol Changes After Data Were Seen}
\label{app:deviations}

The study followed a protocol fixed before collection; changes were logged with time and reason. The following decisions were made after some results were visible. (1) The scoring rule for the browser check (which answers count as a miss of the rendered reference, using the author's notes) was fixed after the answers were read. (2) The primary comparison originally required the realised false-alarm rate on the reviewed sample to be at most 5\%; after interim results showed that no content reader met this, we applied the protocol's statement that thresholds are fixed on validation and report realised false-alarm rates alongside recall, plus a descriptive oracle comparison. Under the original reading the comparison is undefined, because even H-best exceeds 5\%. (3) The protocol's fallback from Qwen3-0.6B to Qwen3-1.7B for M3 was triggered after the 0.6B validation result; both runs are in the release (validation recall 0.60 and 0.66 at 5\% false alarm, against 0.69 for M2). (4) The analyses marked exploratory in the paper (static versus rendered, retention distribution, error analysis, fetch policies, distortion by answer system, content-absent rate of cited URLs) were added while writing, using only data and labels already fixed. (5) After internal audits we added pages-only detector results, the description of H3's dependence on soft-404 sites and the exact review procedure; we also counted cited URLs whose host has no DNS address as dead rather than failed, as the outcome definition requires (20 URLs), and corrected one reviewed label that was typed as a substitute page but marked usable, against the usability rule.

\section{Detectors by Population}
\label{app:bypop}

\begin{table*}[t]
\caption{Recall / false-alarm rate by population on the reviewed test sample, thresholds fixed on validation.}
\label{tab:bypop}
\small
\begin{tabular}{@{}lrrrrrr@{}}
\toprule
 & S1 & S2 & S3 & S4 & S5 & P\\
\midrule
$n$ (unus./us.) & 15/30 & 19/37 & 5/33 & 22/30 & 14/32 & 24/33 \\
\midrule
H-best & 0.93 / 0.03 & 0.79 / 0.05 & 0.80 / 0.03 & 0.95 / 0.10 & 0.86 / 0.03 & 0.79 / 0.09 \\
M1 & 0.60 / 0.00 & 0.37 / 0.08 & 0.60 / 0.09 & 0.59 / 0.10 & 0.50 / 0.06 & 0.71 / 0.00 \\
M2 & 0.93 / 0.10 & 0.84 / 0.16 & 1.00 / 0.03 & 0.95 / 0.13 & 0.93 / 0.09 & 0.88 / 0.27 \\
M3 & 0.80 / 0.00 & 0.68 / 0.27 & 1.00 / 0.03 & 0.86 / 0.13 & 0.79 / 0.16 & 0.88 / 0.18 \\
\bottomrule
\end{tabular}
\end{table*}

Table~\ref{tab:bypop} splits the reviewed-sample results by population. The cells are small (5 to 37 items); M2's false-alarm rate is highest on the cited sources, and M3's on the Tranco top 10k.

\end{document}